# Large-scale cryovolcanic resurfacing on Pluto

Kelsi N. Singer [1✉], Oliver L. White[2], Bernard Schmitt[3], Erika L. Rader[4], Silvia Protopapa[1], William M. Grundy[5], Dale P. Cruikshank[6], Tanguy Bertrand[6,7], Paul M. Schenk[8], William B. McKinnon[9], S. Alan Stern[1], Rajani D. Dhingra[10], Kirby D. Runyon[11], Ross A. Beyer[2], Veronica J. Bray[12], Cristina Dalle Ore[2], John R. Spencer[1], Jeffrey M. Moore[6], Francis Nimmo[13], James T. Keane[10], Leslie A. Young[1], Catherine B. Olkin[1], Tod R. Lauer[14], Harold A. Weaver[11] & Kimberly Ennico-Smith[6]

The New Horizons spacecraft returned images and compositional data showing that terrains on Pluto span a variety of ages, ranging from relatively ancient, heavily cratered areas to very young surfaces with few-to-no impact craters. One of the regions with very few impact craters is dominated by enormous rises with hummocky flanks. Similar features do not exist anywhere else in the imaged solar system. Here we analyze the geomorphology and composition of the features and conclude this region was resurfaced by cryovolcanic processes, of a type and scale so far unique to Pluto. Creation of this terrain requires multiple eruption sites and a large volume of material (>$10^4$ km$^3$) to form what we propose are multiple, several-km-high domes, some of which merge to form more complex planforms. The existence of these massive features suggests Pluto's interior structure and evolution allows for either enhanced retention of heat or more heat overall than was anticipated before New Horizons, which permitted mobilization of water-ice-rich materials late in Pluto's history.

[1] Southwest Research Institute, Boulder, CO 80302, USA. [2] Carl Sagan Center at the SETI Institute, Mountain View, CA 94043, USA. [3] Université Grenoble Alpes, CNRS, IPAG, F-38000 Grenoble, France. [4] University of Idaho, Moscow, ID 83844, USA. [5] Lowell Observatory, Flagstaff, AZ 86001, USA. [6] National Aeronautics and Space Administration (NASA) Ames Research Center, Space Science Division, Moffett Field, CA 94035, USA. [7] LESIA/Observatoire de Paris, PSL, CNRS UMR 8109, University Pierre et Marie Curie, University Paris-Diderot, 5 place Jules Janssen, F-92195 Meudon Cédex, France. [8] Lunar and Planetary Institute, Houston, TX 77058, USA. [9] Department of Earth and Planetary Sciences, Washington University, St. Louis, MO 63130, USA. [10] Jet Propulsion Laboratory, California Institute of Technology, Pasadena, CA 91109, USA. [11] Johns Hopkins University Applied Physics Laboratory, Laurel, MD 20723, USA. [12] University of Arizona, Tucson, AZ 85721, USA. [13] Department of Earth and Planetary Sciences, University of California, Santa Cruz, CA 95064, USA. [14] National Science Foundation National Optical Infrared Astronomy Research Laboratory, Tucson, AZ 26732, USA. ✉email: ksinger@boulder.swri.edu

Pluto's surface has experienced considerable and ongoing resurfacing through both endogenic and exogenic processes[1–3]. Pluto is the largest body in the Kuiper belt with a radius ($R$) of 1188.3 ± 1.6 km[4] and bulk density constraints for a differentiated Pluto indicate the outer ~300 km of Pluto are water-ice-rich overlying a rocky core[5], with a poorly constrained carbonaceous component[6]. Based on this rock abundance, Pluto is expected to have maintained relatively low levels of radiogenic heating (≲5 mW m$^{-2}$) throughout much of its history[7,8]. Pluto's largest moon Charon ($R = 606.0 ±1.0$ km) likely formed through a large, grazing impact with Pluto[9,10]. Models predict the tidal evolution of Pluto and Charon progressed rapidly after the impact, and any tidal heating should have ended very early in their history (<100 Myrs after the impact)[11]. Despite these constraints, modelling suggests a subsurface water-rich ocean could potentially persist into the present on Pluto[8,12–15]. Any ocean is generally predicted to exist 100–200 km or more below the surface of Pluto, at the base of the icy shell[16].

Typical surface temperatures on Pluto are ~35–60 K[17–20], with cooler temperatures for the brighter, volatile-rich surfaces. Pluto's atmospheric surface pressure in 2015 was ~10 μbar[21,22], and no liquid can exist on the surface of Pluto for long owing to this pressure being far below the triple point of the observed ice species ($N_2$, CO, $CH_4$, $NH_3$, $CH_3OH$, and $H_2O$)[23,24]. At these low temperatures pure water ice should generally form an immobile bedrock, as it is also far from its melting temperature of ~273 K. The addition of ammonia or other anti-freeze components (e.g., salts) to the water ice can lower the melting temperature somewhat. The freezing temperature can be depressed by up to ~100 K for high concentrations of ammonia at low pressure e.g.,[25]. Additional antifreeze components could potentially lower the melting temperatures even further[26], but the surface temperatures on Pluto are so cold and the atmospheric pressure so low that freezing of a fluid on the surface would still occur on relatively short geologic timescales[23]. On Pluto's surface, nitrogen ice ($N_2$) is much closer to its melting temperature (63 K) than water ice, and can flow or viscously relax over relatively short timescales[27,28]. Volatile ices ($N_2$, CO, $CH_4$) also play a role in resurfacing areas of Pluto through sublimation, physical erosion, and/or deposition/mantling[24,29–32].

Here we show that the potential icy volcanic (or cryovolcanic) constructs and their surrounding terrain discussed here (Fig. 1) have many morphological traits that are distinct from any other area on Pluto. These geologic features do not appear to be formed predominantly by erosion nor do they appear to be constructed primarily of volatile ices. Here we refer to cryovolcanism as the collection of processes that cause mobile subsurface material to extrude onto the surface and either partially or fully resurface the existing terrain. We propose a large volume of material has erupted from multiple sources (and likely in more than one episode over time) to form the many large domes and rises found in this region.

## Results

**Morphological characteristics**. The region of putative cryovolcanic terrains discussed here lies to the southwest of the Sputnik Planitia ice sheet (Supplementary Fig. 1), which fills an ~1000-km-diameter ancient impact basin[2,33–36]. The most prominent and largest-scale structures in the cryovolcanic region are large rises or mounds of material separated by broad depressions (Figs. 1–3 and Supplementary Fig. 1). The configuration of the large rises gives the impression of annular features with deep central depressions in two cases. These features are named Wright and Piccard Montes. However, further inspection suggests these features may or may not be annular, and instead may simply have arisen from the merging of several adjacent rises (discussed below). The main topographic rise of Wright Mons (Fig. 1) stands ~4–5 km high (relative to the lower areas of surrounding terrain) and spans ~150 km, and Piccard Mons (Supplementary Fig. 2) is ~7 km high at its tallest points and ~225 km wide. The inferred volume of the main topographic rise of Wright Mons alone is ~2.4 x $10^4$ km$^3$ (similar to the volume of Mauna Loa, see Supplementary Note 8).

Wright Mons was imaged in sunlight but was also located near the terminator (transition from night to day) during New Horizons closest approach. Thus, the incoming sunlight is at a fairly low angle close to the surface (<30° elevation angle), and uni-directional (from the north-west) and this creates an effect where features perpendicular to the lighting direction (roughly NE-SW) are emphasized. Here we focus on features that can be verified with topographic data and can be seen in multiple image datasets with different lighting geometries. Piccard Mons had rotated past the terminator by the time New Horizons conducted its highest resolution imaging. Reflected light from high-altitude hazes in Pluto's atmosphere allowed for some higher-resolution imaging of Piccard Mons past the terminator (although at a lower signal-to-noise ratio than the sunlight regions)[2,36]. Other large rises lie between Wright and Piccard Montes (here referred to as the medial montes region), and seem to be connected to Wright and Piccard with no sharp transition in surface morphology, and, in some areas, no sharp transition in elevation (e.g., area labelled "B" and "C" in Fig. 1; also see Supplementary Fig. 3).

The flanks of Wright Mons and much of the surrounding terrain including the nearby large rises exhibit an undulatory/hummocky texture that varies in wavelength/scale from a few to ~20 km across, with the most common widths between 6–12 km across (Fig. 1c, Supplementary Figs. 4–5). The hummocky terrain has either flat or gently rounded tops and is irregular in planform; most are interconnected on one or more sides and not individual mounds (although we still use the word hummocky here as a general reference to the type of texture). The lows between hummocks also vary, as some are narrow compared to the swells (with v-shaped profiles) and some are similar in width to the swells (more U-shaped profiles). The trough depths/hummock heights also vary and are typically between a few hundred meters and 1 km (Supplementary Figs. 4–5). Although the oblique lighting makes it appear as if the northern flank is smoother (not as hummocky), the topography shows that the hummocky terrain exists on all flanks of Wright Mons (Fig. 1b). The hummocky texture does not appear to have a preferential orientation (relative to the central depression or otherwise). On yet a smaller scale, boulders, blocks, slabs, or ridges with a horizontal scale of ~1–2 km are superimposed on the hummocks (Fig. 1c). These smallest-scale features are only 3–10 pixels across thus are difficult to characterize.

The large-scale slopes across the broad flanks of Wright Mons are ~3–5° (reaching 10° in some locations). The central depression of Wright Mons is ~40–50 km across, and extends down to approximately the level of the surrounding terrain or slightly below (Fig. 1d), making it ~4 km deep on average. The central depression of Piccard Mons is even larger in size and has a more rounded or "U-shaped" profile[36]. This central depression is dissimilar to calderas on terrestrial or martian volcanos, as it occupies about 1/3rd of the overall width of the features, is very deep (i.e., its depth is equal to the height of Wright Mons) with a quasi-conical shape (i.e., it is not a smaller depression at the summit of a large shield, dome, or cone), and no traditional collapse terraces or similar structures are apparent (Supplementary Fig. 6d, e). The central depression walls are also lumpy in appearance, similar to the outer flanks, with typical large-scale slopes of ~a-few-to-10° (up to ~20° in a few locations). Topographic profiles show the northern flank of Wright, its southern flank, and the adjacent swell of the medial montes region all have similar topographic profiles (Fig. 1d).

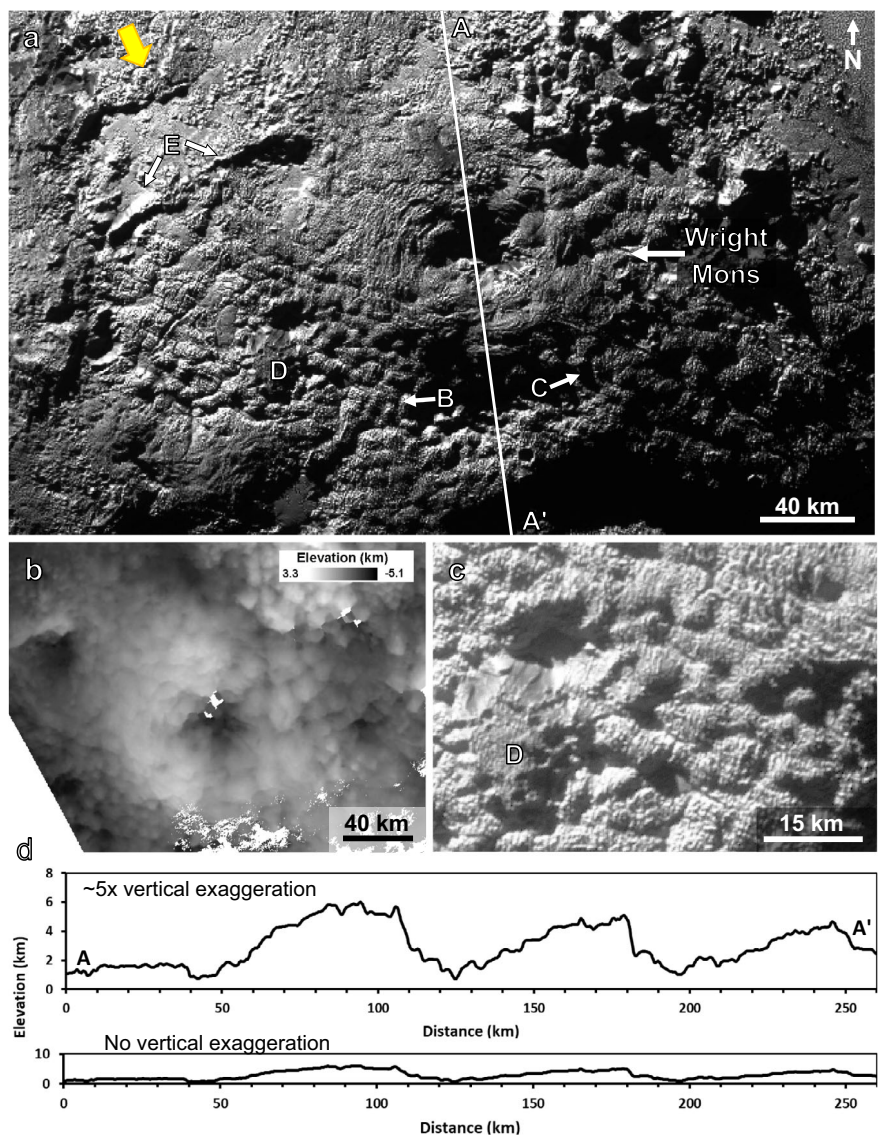

**Fig. 1 Features of Wright Mons and the surrounding terrain. a** Wright Mons region with features labelled (see text), **b**, high-resolution topography for Wright Mons[36], **c**, zoom of region with smaller dome named Coleman Mons (label "D"; also see Fig. 4), undulating, hummocky terrain on the flanks of Wright Mons and the superposed smaller-scale (1–2 km) ridges or boulders, **d**, topographic profile of Wright Mons and adjacent rise as shown by the line A to A' in panel a. All images are from the new Horizons observation PEMV_P_MVIC_LORRI_CA (315 m px$^{-1}$; see Supplementary Table 1) on a simple cylindrical projection. The large arrow in the upper left of panel a indicates the direction of incoming sunlight and is repeated in subsequent figures. All figures in the main text and supplement are shown with north up. The longitude and latitude extents of the image are as follows: panel a ~163–182°E and ~16–28°S; panel b ~166–177°E and ~17–24°S; panel c ~167–171°E and ~22–25°S.

If the central depressions in Wright and Piccard Montes were formed entirely due to the collapse of the summits of formerly mound-shaped or conical edifices, this would represent the removal of >50% of the edifices' volumes, a vast fraction. A comparison with shield volcanos on Earth and Mars (Supplementary Fig. 7) highlights how different the shape of the features on Pluto are and how atypical the central depression of Wright Mons would be if it were a collapse feature. A few other irregular depressions with steeper walls of various sizes (a few to 30 km across and a few hundred to a few km deep) are scattered throughout the terrain; most cavi do not appear to be impact craters because of their lack of both circularity and raised rims. Some of the depressions have sharp flat edges that may represent fault faces where collapse played a role in forming the depression, whereas others may be formed simply by the rise of material around them. These could potentially represent vent sites, but there are no clear indications of the flow of material from them.

There are no obvious indicators of flow directionality or locations of effusive centers. Any distinct flow fronts, streamlines, levees, or fractures/vent locations that may have formed are not evident at the resolution of New Horizons images (the best images in this area range from 234–315 m px$^{-1}$), or may have been degraded due to post-formation processes. However, there are some indications that multiple resurfacing events or emplacement episodes may have occurred (see methods). There are also no obvious indicators of explosive volcanism[37], such as ballistic fall deposit patterns (either radial or directional), or steeper cones. The full extent of the resurfaced terrains is not known, as these terrains continue southward until they are no longer visible in the haze-light[36]. The scarcity of craters on Wright Mons indicates a relatively young age, with a previously determined upper limit of ~1–2 Ga[38]. Given uncertainties in the impactor flux onto Pluto, and small number statistics, the crater retention age does not present a strong constraint, and many features in this area could be considerably younger.

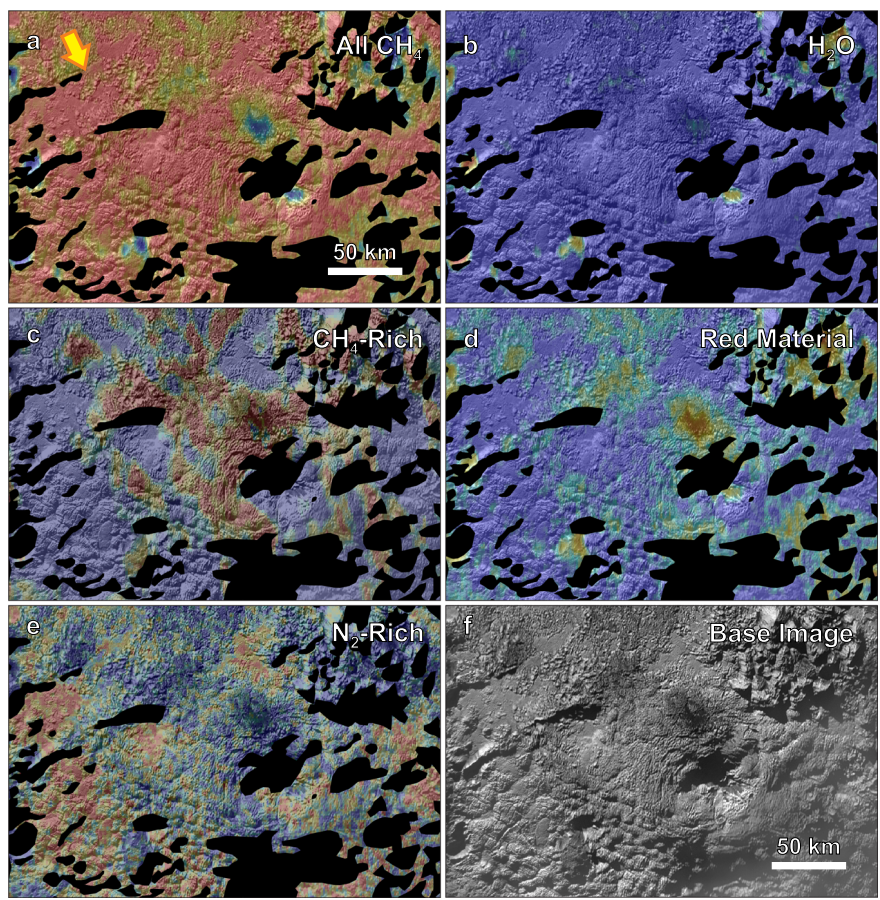

**Fig. 2 Surface composition of the Wright Mons region.** Color-scale composition maps from ref. [41] derived from LEISA spectral data (see Supplementary Table 1) overlain on a greyscale panchromatic basemap. In all panels, redder colors indicate a greater absorption/band depth or a greater spectral index, indicating a stronger presence of the material. The range of values for each index is given in this caption. **a**, methane ($CH_4$) band depth (with values from −0.08 to 0.43) wherever this molecule appears (as $CH_4$-rich ice or in $N_2$-rich ice), **b**, water ice spectral index ($H_2O$) (with values from −1.14 to 0.93), **c**, areas where methane dominates over nitrogen ice ($CH_4$-rich only, found generally at higher elevations; with values of the '$CH_4$ bands position index' from 39 to 47; see ref. [41] their Fig. 22), **d**, an organic dark, red (in the visible) material index (with values from −1.94 to −0.18), **e**, areas where nitrogen ice ($N_2$) dominates over methane (with values of the '$CH_4$ bands position index' from 42 to 32), and **f**, the panchromatic basemap alone. Please note that the shadowed regions have been excluded here because the low light levels make them difficult to accurately characterize with the LEISA data. The data is shown in a simple cylindrical map projection. The longitude and latitude extents of the image are ~164–179°E and ~16–25°S. The large arrow in the upper left indicates the approximate direction of the incoming light. Methane dominates much of the region but water ice (mixed with dark red material) is apparent on dark/low albedo patches, which are presumably warmer areas where methane ice is not stable.

**Compositional constraints**. Methane, nitrogen, and water ice are all observed to exist in high-volume, concentrated deposits on the surface of Pluto[24]. Thus, we consider whether these materials could make up the bulk of the cryovolcanic units that make up Wright and Piccard Mons and their surrounding terrains, based on both observations by New Horizons and what is known about the characteristics of the materials.

The Linear Etalon Imaging Spectral Array (LEISA) instrument on New Horizons[39] acquired infrared spectroscopic data informative of Pluto's composition[24,40,41]. The volatile ices $N_2$, CO, and $CH_4$ form complex multi-phase systems as $N_2$-rich and $CH_4$-rich mixtures across much of the surface of Pluto (Fig. 2) because they sublimate and redeposit following seasonal cycles (Pluto's year is 248 Earth years) or the longer multi-million-year obliquity/precession cycles e.g.,[24,42–45]. In darker, low-albedo, warmer areas across Pluto, volatile ices do not deposit (or are not stable) and the spectral signatures of the non-volatile water ice "bedrock" and a dark organic material can be observed instead (e.g., in the dark equatorial band on Pluto). This pattern can be seen in the Wright Mons region as well (Fig. 2). As the methane spectral signature becomes weaker on the few dark surface areas around Wright Mons, the water ice and red material signal becomes stronger (Fig. 2a–d). Methane-rich ice is also more prevalent at higher elevation (Fig. 2c). This indicates that the methane is likely a thin surface layer deposited out of the atmosphere[46], and the bulk of Wright Mons and the other large topographic features in the area are not necessarily composed of methane. Additionally, the Wright Mons region exhibits a very different surface texture than that of the "bladed terrain" on Pluto (Supplementary Fig. 8), which is thought to form by condensation and sublimation of thick methane deposits[46–48].

The spectral signature of nitrogen ice is also found across the Wright Mons region (Fig. 2e), appearing as both smaller, smooth, nitrogen-rich ice patches likely ponded in local lows, and also across the scene in a distribution similar to the thin methane deposits at lower elevation. However, as previously mentioned, larger volumes of nitrogen-rich ice cannot maintain tall topographic relief at Pluto's surface conditions[27,34].

Thus, for the remainder of the paper, we will explore ideas for forming the terrain in the Wright Mons region out of predominantly water ice, with the potential for other materials to be mixed in that may have aided in the deposition or further sculpting of the terrain over time. Ammonia or an ammoniated compound has been detected near extensional fractures (~130°E, 10°N) on Pluto where cryofluid eruption may have brought it to the surface in a thin deposit[23,49]. No clear signature of ammonia is observed in the region described here (Dalle Ore and Cruikshank personal communication), although it could be obscured by the methane signature. The dark material itself is mostly thought to be a class of materials called tholins[24,33,50,51], which are disordered and insoluble carbon-rich macromolecular materials resulting from the energetic processing of hydrocarbons (e.g., $CH_4$) and other molecules containing nitrogen, carbon, and/or oxygen. The majority of the darker deposits in this region occur on north-facing slopes, which can be explained by insolation patterns[46].

In addition to the spectral data, color observations (Fig. 3) from the New Horizons Multispectral and Visible Imaging Camera (MVIC) are helpful for distinguishing compositional differences across terrains[39,51,52]. The dark material on some north-facing slopes has a very strong red spectral slope (as seen in the brighter, redder areas of the enhanced color image in Fig. 3). There are also more subtle albedo and redness variations across the Wright Mons region. For example, much of Wright Mons has a slight red color, whereas the terrain just to its north is redder. The morphological transition from the Wright Mons region to the large plateau to the west (transition region labelled "A" in Fig. 3) is also reflected in a color transition (from redder to less red). Although it is difficult to determine age relationships corresponding to the variation in albedo or redness of terrains, the existence of albedo variation may be indicating these regions have been emplaced at different times, from varying source reservoirs, or from variations on the extrusion process.

**Material emplacement hypotheses**. The constraints above suggest that these voluminous, potentially water-ice-rich structures were emplaced on the surface of Pluto in the later part of its history. Our new analysis concurs with the previous discussion that the features are likely constructional[2,3,36,53] from the cryovolcanic emplacement of material on the surface, and are not erosional remnants or features formed purely from uplift from below. The three main lines of evidence that combine to suggest constructional features are: (1) this enormous area of resurfaced terrain has a paucity of craters (with no unambiguous examples), implying the formation event(s) reset the surface, (2) the hummocky morphology of this region is found on both the flanks and crests of rises as well as on lower, flatter terrain, and is dissimilar to the appearance of terrains scoured by glacial erosion or resurfaced by volatile sublimation-erosion found elsewhere on Pluto, and (3) these features lie well above their surrounding terrain in a variable pattern of highs and low, and thus cannot realistically be erosional remnants.

Given the spatially associated nature and morphological and topographical similarity of all of the large rises in the region, we put forward a new hypothesis: Wright Mons (and similarly for Piccard Mons) may be comprised of multiple, separate rises that have merged in some areas but not others, and that share the same formation mechanism as all of the other large rises and domes in the area. This is a departure from previous studies that considered Wright Mons more similar to single, coherent edifices with a central caldera and the other large rises as a separate kind of feature. This hypothesis is also consistent with the base of the central depression in Wright Mons sitting at a similar elevation as the surrounding terrain (although Piccard's central depression is deeper than much of the surrounding terrain).

The smaller dome-like feature, Coleman Mons (labelled "D" in Fig. 1; Fig. 4), may represent an example of how the material is emplaced in this region. If this dome has a central, main source vent, then a dome of ~25 km in diameter and ~1.5 km high would imply a basal yield strength of ~$6 \times 10^4$ Pa in the dome growth model of Bridges and Fink[54] (see methods). This yield strength value is consistent with some measures of ductile strength of mobile water or ammonia-water ice (~$10^4$–$10^5$ Pa)[55,56], which is to be expected if these features are formed of somewhat more mobile ice.

The hummocky/ropey nature of the flanks of Wright Mons and the surrounding terrain are suggestive of viscous flow of either slushy or solid-state but still mobile material. We investigated three hypotheses to create the undulating/hummocky texture: (1) creation of individual small volcanic domes (first proposed in[36]), (2) viscous extrusion of rapidly cooled lavas analogous to pillow lavas, (3) compression of viscous material with a frozen skin analogous to pahoehoe, viscous pressure ridges, or funiscular terrain on Enceladus[57]. We also consider the potential role of fractures in the area to control extrusion patterns or erosion.

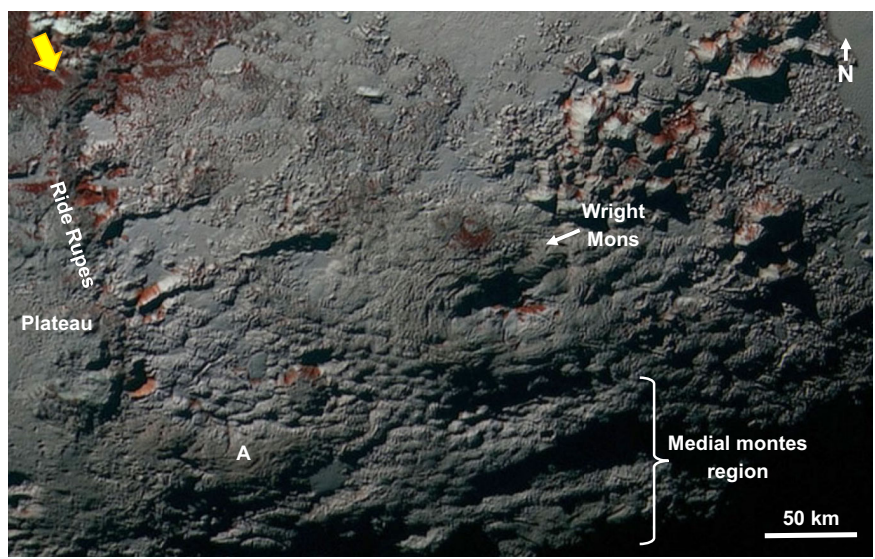

**Fig. 3 Color information for the Wright Mons region.** Darker/lower albedo, redder patches exist primarily on north-facing slopes but there are also more subtle differences in albedo and redness across the region. The region labelled "A" represents a redder unit transition to less red units at lower elevation (described in the text and methods). From the New Horizons observation PEMV_P_Color2 (~660 m px$^{-1}$) shown in the original image geometry. The longitude and latitude extents of the image are ~160–182°E and ~13–31°S.

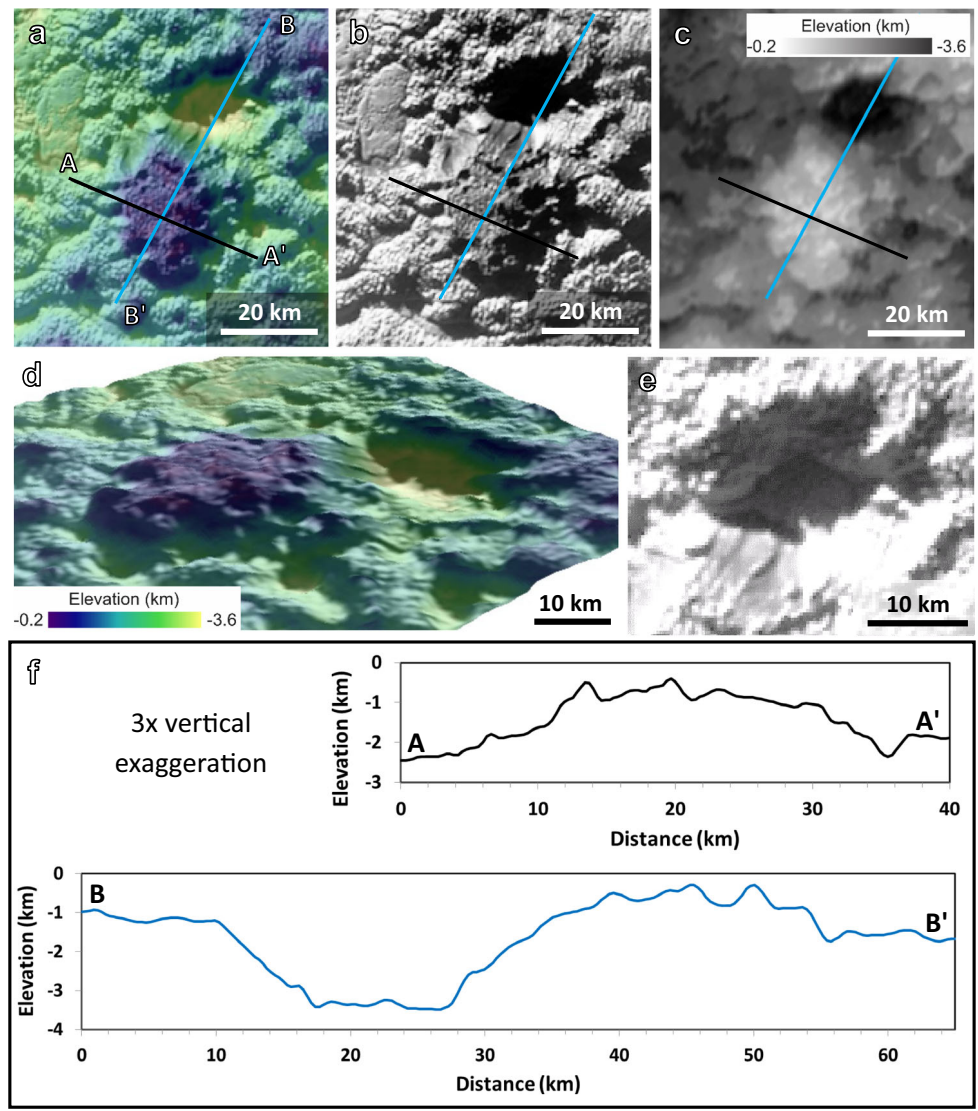

**Fig. 4 Smaller dome-like feature named Coleman mons. a** Topography[36] overlain on base image of feature, **b**, base image alone, **c**, topography alone, **d**, perspective view of dome and pit with no vertical exaggeration, **e**, view inside the pit, **f**, topographic profiles as shown in panels a-c, with ~3× vertical exaggeration. All images from the New Horizons observation PEMV_P_MVIC_LORRI_CA (315 m px$^{-1}$; see Supplementary Table 1), shown in a simple cylindrical projection. The location of this feature is indicated by the letter "D" in Fig. 1.

For both the creation of individual domes or a process similar to pillow lava formation, subsurface source material would need to be extruded at a similar rate and for a similar duration across both the plains and the flanks/tops of the large rises to create similarly sized hummocks (for some details of potential cryomagma extrusion on other worlds see e.g.,[58–61]). Such uniform extrusion over such a diverse terrain seems unlikely. It is possible that the hummocky resurfacing occurred first, and was subsequently uplifted to form the large rises. This would imply an enormous volume of intrusion under the hummocky surface. If the hummocks are contractional features, a rough estimate of the thickness of the high-viscosity layer required to achieve a "folding wavelength" similar to the hummocks in diameter is 8–13 km, which makes this mechanism unrealistic (see methods). Additionally, it is not clear what could cause compression.

Relatively large, deep fractures would presumably be needed to act as conduits for the ascent of subsurface mobile material in any of these scenarios (or a fracture network could also act as a mechanical filter to control hummock size/spacing). Although there are fractures across much of Pluto[2,3,36], there are not many obvious large fractures in the Wright Mons region. The very large scarp (Ride Rupes; Fig. 3 and Supplementary Fig. 1) separating the Wright region from the plateau to the west, and one other scarp (Fig. 1, label "E") are the only visible indications of possible deep fracturing in the Wright Mons region. The extrusive process may have covered other deep fractures.

## Discussion

The scenarios described above illustrate how canonical models of emplacement (derived primarily from terrestrial studies) may not be directly applicable to Pluto. The geologic features in the Wright Mons region are morphologically unlike any other regions on Pluto and also have very few similarities to most terrains on other bodies in the solar system. The lack of indications of source vent regions or directionality of material movement makes it difficult to positively determine the mechanism of material emplacement on the surface. However, we found through detailed examination of all New Horizons imaging and composition data available for the Wright Mons region that the many, large, morphologically complex cryovolcanic constructs are consistent with formation from multiple subsurface sources where the sources are below the constructs. This scenario allows for a consistent formation mechanism for all of the large rises and depressions—where some are domical or annular and others are complex shapes—through the merging of different rises. It also avoids the need for an enormous amount of collapse to explain the giant depressions.

Given the low expected heat fluxes from Pluto's interior, and Pluto's cold surface temperatures (both topics discussed in the introduction), mobilizing material primarily made up of water ice is thermally challenging. However, the relative youth of the terrains implies that some heat must be available to emplace these features late in Pluto's history. Multiple, massive water-ice cryovolcanic constructs present new pieces of information towards understanding Pluto's thermal history, which complement other information from young areas on Pluto made up of volatile ices (e.g., Sputnik Planitia), and other small-volume features that have been proposed as effusions of ammonia water[23,62]. Perhaps the stratigraphic arrangement of the interior structure has stored internal heat generated from the rocky core that was later released (e.g., the clathrate layer proposed by ref. [14]).

The range of cryovolcanic features found across the solar system is diverse. With the different conditions and surface materials present at Pluto, it is quite possible that any material movement onto the surface may not resemble that of other bodies. The extrusion of icy material onto the surface of a body with extremely low temperatures, low atmospheric pressure, low gravity, and the abundance of the volatile ices found on Pluto's surface make it unique among the visited places in the solar system.

## Methods

**Pluto topography from stereogrammetry.** Multiple stereo image pairs were available for the creation of several digital elevation models for the hemisphere of Pluto visible at encounter. These models were integrated into one final topography map product[36]. The New Horizons images and the final integrated topography map product is available from the Planetary Data Systems Small Bodies Node located at https://pds-smallbodies.astro.umd.edu/data_sb/missions/newhorizons/index.shtml, with the topography map in the subdirectory https://pds-smallbodies.astro.umd.edu/holdings/nh-p_psa-lorri_mvic-5-geophys-v1.0/data/dtm/. The production of this map is described in ref. [36], and we also provide some additional details here. Image registration and creation of stereo pairs was completes using United State Geologic Survey planetary image processing software (https://doi.org/10.5281/zenodo.3962369). To estimate a feature's height, its displacement or parallax is first determined using scene recognition. For this, a 3x3 pixel box size is used, thus the effective horizontal ground pixel scale of the resulting topography is ~3 times the pixel scale of the lowest resolution image in the stereo pair. The standard photogrammetric parallax equations[63] are then used to determine the distance to points on the body. Topography produced through stereogrammetry were also cross-checked with feature heights from shadow measurements where applicable and were found to be indistinguishable. The portion of the topographic map for Wright Mons and its surroundings was created with the following stereo pair: PELR_P_LEISA_HIRES image sequence (240 m px$^{-1}$) and PEMV_P_MVIC_LORRI_CA image scan (315 m px$^{-1}$), with an effective horizontal resolution of 945 m px$^{-1}$ and vertical precision of ~90 m (see Supplementary Table 1 and also see Table 1 in ref. [36]). For the wider area, an additional stereo pair filled in additional terrain: PEMV_P_MVIC_LORRI_CA image scan (315 m px$^{-1}$) and PEMV_P_MPAN1 image scan (480 m px$^{-1}$), with an effective horizontal resolution of 1440 m px$^{-1}$ and vertical precision of ~230 m. See Fig. 7 in ref. [36] and Supplementary Fig. 1 in ref. [64] for image sequence extents displayed on the Pluto base map.

**Possible evidence for multiple episodes of emplacement.** Several features of the Wright and Piccard region may point to the terrain being created in more than one event. We describe four features here. (1) Terrain to the north of Wright mons (Supplementary Fig. 6a) has a somewhat similar small-scale texture (1–2 km boulders/ridges), although it lacks obvious mid-sized (~8–12 km) hummocks. This northern surrounding terrain is somewhat darker and overprinted by what appear to be a few small craters. (2) Lower elevation plains directly to the west of Wright Mons (Supplementary Fig. 6b, c) have a similar undulating/hummocky appearance to Wright Mons, but are also superposed by an intersecting fracture set. These fractures mostly appear fairly shallow (as if they do not cut all the way through the hummocks) but a few are deeper. These more modified terrains may represent an earlier episode of the process that created Wright Mons and the other large rises, that have subsequently been more cratered or tectonized. Additional there are several possible examples of superposed flows or episodes of terrain emplacement. (3) Coleman Mons (Fig. 4 and described in the next section) may represent an example of a separate emplacement event on the surface. And finally, (4) at the southern extent of Ride Rupes, the terrain between the Wright region and the large plateau farther to the west are connected by a gradual transition in elevation, albedo, color, and morphology (Fig. 3, labelled "A", and Supplementary Fig. 6f): from hummocky to less hummocky to pitted moving east to west. Although there are no clear contacts, the higher elevation materials are somewhat darker and may superpose the lower elevation brighter units and may indicate these were separate emplacement events. Alternatively, the material may have been emplaced at the same time but later events more heavily modified the material at higher elevation.

**Dome model for Coleman Mons.** We work with the hypothesis that material may extrude from below Coleman Mons (Figs. 1c and 4), making this a small dome-like structure. The dome is made out of darker material and sits ~1 km above the tops of the surrounding hummocks. It is lumpy but not as clearly hummocky as some of the surrounding terrain. The darker material seems to cover parts of the surrounding hummocks (without disrupting them), and represents one of the few more distinct contacts in the area. Coleman Mons sits next to a depression (that reaches several km below the surface) but Coleman Mons is not obviously associated with the depression (Fig. 4e).

If this feature represents a smaller dome, it could be indicative of the mode of emplacement in the Wright Mons region. The rheology of the material being extruded can be related to the dome shape[54]. The dome is somewhat oblong in planform, with long and short axes of ~30 and 20 km, respectively. The hummocky nature of the terrain around the dome makes heights less straightforward to measure than in most terrestrial examples, but measurements range from 1 to 2.5 km around the dome. Using an average/typical diameter and height of 25 km and 1.5 km, respectively, gives an aspect ratio ($A$ = height/diameter) of ~0.06 (with a range of ~0.125 to 0.03 for the range of diameter and height measurements). The aspect ratio can be related to the dome geometry and material parameters through $A \approx V^{-0.2} \tau_{base}^{0.6} \rho^{-0.6} g^{-0.6}$ e.g.,[54,65], where $V$ is the volume of a circular dome, $\tau_{base}$ is the shear strength at the base of the expanding dome, $\rho$ is the lava density (we used 920 kg m$^{-3}$ for this example, as a lower limit for cold, pure, water ice), and $g$ is surface gravity (0.62 m s$^{-2}$ for Pluto). Because the basal shear stress during flow for these materials is not well known, we use the measured aspect ratio to estimate what shear stresses would match the observed dome geometry. Bridges and Fink[54] argue that the basal shear strength will be equal to the yield strength for low strain rates typical of growing domes (at least in terrestrial examples). For an intermediate aspect ratio of 0.06 for the dome on Pluto, the estimated basal shear stress or yield stress is $6.4 \times 10^4$ Pa, with a range of $2.4 \times 10^4$ to $2.1 \times 10^5$ Pa for the extreme range of aspect ratios. A similar equation for a dome with a roughly parabolic cross-section (although this dome is somewhat more flat-topped) produces a similar estimate of basal Bingham yield stress, $5.1 \times 10^4$ Pa for the average dome dimensions[66].

Relevant laboratory measurements of Bingham yield strength for pure water ice are not, to our knowledge, available. Measurements for ductile strength of both water ice and ammonia-water ice slurries under a confining pressure of 50 MPa (higher than expected for the features on Pluto's surface) are in the range of $10^4$–$10^5$ Pa at temperatures where the ice is still mobile (~140–170 K for ammonia water ice)[55,56]. These are not the same conditions as on Pluto's surface, but the ice would presumably still need to be at temperatures where it was mobile in the interior of the flows. The yield strengths calculated for Coleman Mons are also in the range of estimated values for terrestrial and lunar values basaltic and rhyolitic values ($10^3$–$10^5$ Pa)[54].

Several ~3–19-km-diameter domes on Europa have been modeled as cryovolcanic emplacements[58,60,67], however, they are considerably less tall features (30–100 m) than Coleman Mons or the other large rises in this area of Pluto. The Europa examples also have somewhat smoother surfaces to the extruded material (although some also have rafted ice blocks) and more regular dome-like shapes. Although the temperatures and surface gravity ($g$) on Europa are not as low as on Pluto (Europa: 100 K for the aforementioned models and $g$ = 1.315 m s$^{-1}$; Pluto: average 40 K and $g$ = 0.62 m s$^{-1}$), more complex volcanic extrusion or dome formation modeling such as investigating possible ascent mechanisms, cooling rates, or dome relaxation[58,60,68] may be fruitful avenues of future research.

**Funiscular terrain analogy**. The hummocky terrain and the smaller scale boulders and/or ridges superimposed on them bear some resemblance to areas of the funiscular terrain found between the tiger stripes of Enceladus (Supplementary Fig. 9)[69], although funiscular terrain is often more linear and has a smaller width (closer to 1 km in wavelength) and amplitude (~0.5 to 1 km height). A leading hypothesis for forming funiscular terrain is through contractional folding of a thin frozen "lithosphere" overlying more viscous material[57,70], akin to formation of pahoehoe textures on Earth. In this case the tiger stripes are in extension resulting in compression between them. On Enceladus, high heat flows and a warm effective surface temperature are needed in the modelled conditions to keep the surface layer thin enough to produce the observed features[57]. Bland, McKinnon[57] suggested the effective surface temperature between the stripes could be higher (possibly 70 K to 186 K) than the measured temperatures of the optical surface (55 K) due to insulation from fractures, porosity, and fallback of fine-grained plume material.

The surface temperatures expected for Pluto are on average ~40 K (described in the main text). Following the folding model of Fink[71], also described in Barr and Preuss[70], the thickness of a high viscosity layer ($H$) needed to create a given dominant folding wavelength ($L_D$), is given by $H \approx (L_D/28)*\ln(R)$, and $\ln(R) \approx (Q^*\Delta T/R_G T_i^2)$. The variables in ln(R) are as follows: $Q^*$ is the rheological activation energy, $\Delta T = (T_i - T_s)$ or the interior temperature minus the surface temperature, and $R_G$ is the gas constant (8.314 J mol$^{-1}$ K$^{-1}$). We use $Q^* = 60$ kJ mol$^{-1}$ for ductile water-ice[72], $L_D = 10$ km as the average wavelength of the hummocky terrain, and vary the internal temperature ($T_i$) from 150 to 273 K as a wide range of possible temperatures for materials ranging from mobile ice with large amounts of antifreeze to liquid water. This yields a thickness range for the high viscosity upper folding layer of $H \sim 8$–13 km, which is unrealistically large for the scale of the features.

Additionally, for Wright (or Piccard) Mons on Pluto, the lack of distinct flow fronts or source regions means that it is not clear what could cause compression to create a folded surface. If the material flowed downhill while continually freezing at the flow front, gravity and pressure from continually erupted material could serve that role. However, the hummocky terrain also occurs on flatter areas. It is additionally not clear if large volumes of material are erupted at a given time, in order to create long flows with a relatively consistent wavelength.

### Data availability
All New Horizons image and topographic data used in this study are publicly available in NASA's Planetary Data System archive (Small Bodies Node: https://pds-smallbodies.astro.umd.edu/).

## Acknowledgements

We thank NASA's New Horizons mission for funding (grant numbers NASW-02008 and NAS5-97271/TaskOrder30), and the New Horizons team for their hard work leading to a successful Pluto system flyby and subsequent return of the data. KNS additionally thanks Dr. Michael Bland for helpful discussions. BS acknowledges the Centre National d'Etudes Spatiales (CNES) for its financial support through its "Système Solaire" program. A portion of this research was carried out at the Jet Propulsion Laboratory, California Institute of Technology, under a contract with the National Aeronautics and Space Administration (80NM0018D0004).

## Author contributions
K.N.S. conducted the research, and wrote most of the paper. O.L.W. contributed ideas throughout the paper and to the writing. B.S. provided data in Fig. 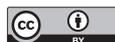 2 and discussed implications and contributed to the writing. E.L.R. contributed ideas and to the model calculations. S.P., W.M.G., D.C., T.B., P.M.S., W.B.M., S.A.S., R.D., K.D.R., R.A.B., and V.J.B. contributed ideas and to the writing. C.D.O., J.R.S., J.M.M., F.N., and J.T.K. discussed the results and contributed ideas. L.A.Y., C.A.B., T.R.L., H.A.W., and K.S.E. contributed to the planning and successful data collection and analysis of the New Horizons mission that made these results possible.

## Competing interests
The authors declare no competing interests.

## Additional information
**Supplementary information** The online version contains supplementary material available at https://doi.org/10.1038/s41467-022-29056-3.

**Correspondence** and requests for materials should be addressed to Kelsi N. Singer.

**Peer review information** *Nature Communications* thanks Caitlin Ahrens, Kate Craft and the anonymous reviewer(s) for their contribution to the peer review of this work.

**Reprints and permission information** is available at http://www.nature.com/reprints

**Publisher's note** Springer Nature remains neutral with regard to jurisdictional claims in published maps and institutional affiliations.

# Supplement: Large-scale cryovolcanic resurfacing on Pluto

*Singer et al., Nature Communications*

**Supplementary Note 1: New Horizons Image Data**

Wright and Piccard were viewed in several datasets listed in Supplementary Table 1. The best combination of lighting, resolution, and signal-to-noise comes from the ~315 m px$^{-1}$ Multi-spectral Visible Imaging Camera (MVIC) scan taken on closest approach. We used all available imaging of the Wright and Piccard region with different lighting geometries and also topographic products[1] when examining the morphology of this region. We also used simulated hillshades (shaded relief maps) to check for biases introduced by specific lighting conditions.

**Supplementary Table 1: High-resolution New Horizons image data for Pluto**

| Request ID[*] | Instrument[†] | Instrument Mode | Pixel Scale [m px$^{-1}$] | Mosaic size or Scan[‡] | Exposure or Scan rate |
|---|---|---|---|---|---|
| PELR_P_LORRI | LORRI | 1×1 | 850 ± 30 | 4×5 | 150 ms |
| PELR_P_LEISA_HIRES[‡] | LORRI | 1×1 | 234 ± 13 | 1×12 | 50 ms |
| PELR_P_MPAN_1[‡] | LORRI | 1×1 | 117 ± 2 | 1×27 | 10 ms |
| PELR_P_ MVIC_LORRI_CA[‡] | LORRI | 1×1 | 76 | 1×35 | 10 ms |
| PEMV_P_MPAN1 | MVIC | Pan TDI 1 | 480 ± 5 | Scan | 1600 μrad s$^{-1}$ |
| PEMV_P_ MVIC_LORRI_CA | MVIC | Pan TDI 2 | 315 ± 8 | Scan | 1000 μrad s$^{-1}$ |
| PEMV_P_Color2 | MVIC | Color | 660 | Scan | 1045.5 μrad s$^{-1}$ |
| PELE_01_P_LEISA_Alice_2a | LEISA | | ~7,100 | Scan | 105.1 μrad s$^{-1}$ |
| PELE_01_P_LEISA_Alice_2b | LEISA | | ~6,400 | Scan | 105.1 μrad s$^{-1}$ |
| PELE_01_P_LEISA_Hires | LEISA | | ~3,100 | Scan | 105.1 μrad s$^{-1}$ |

[*]The Request ID are unique identifiers for each observation stored at NASA's Planetary Data System (https://pds-smallbodies.astro.umd.edu/data_sb/missions/newhorizons/index.shtml).
[†]For more information about the New Horizons instruments, please see [2,3]
[‡]Given in number of instrument field-of-view footprints across the mosaic.
Note: All observations listed here were taken within a short time of each other near the closest approach of the New Horizons spacecraft when the sub solar point was at ~128.3°E, 51.6°N.

**Supplementary Note 2: Regional Context of the Wright and Piccard Montes Region**

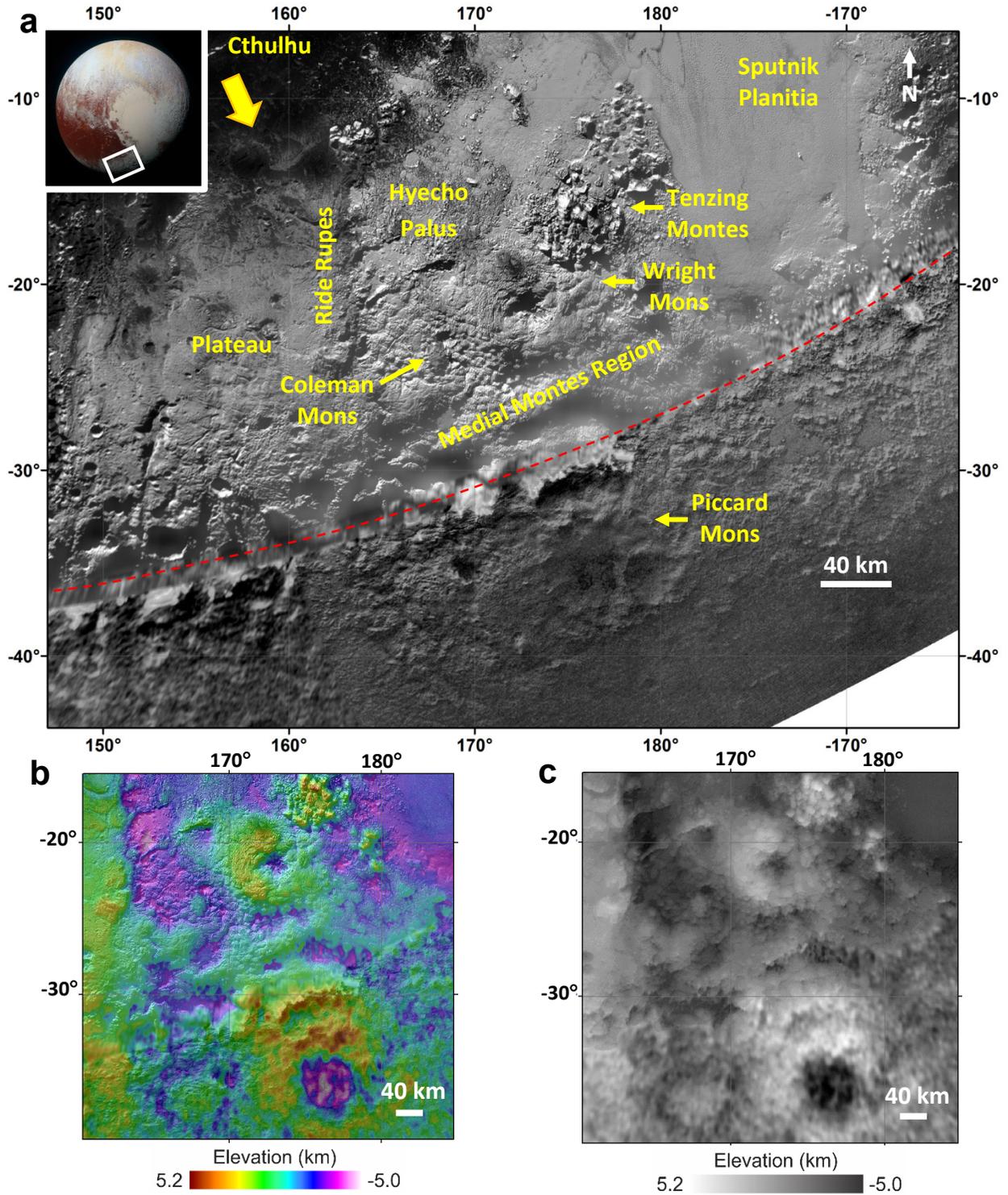

**Supplementary Figure 1 | Overview of putative cryovolcanic terrains. a,** Panchromatic basemap mosaic (image resolutions ranging from ~235-480 m px$^{-1}$; see Supplementary Table 1); dashed red curve indicates the transition from directly sunlit terrain to haze-lit terrain, **b,** color

topography overlain on basemap [1], and **c,** greyscale topography alone. Simple cylindrical projection. All images in the supplement are shown with north up, and the lighting direction is the same as indicated here with the large arrow at the upper left unless otherwise indicated.

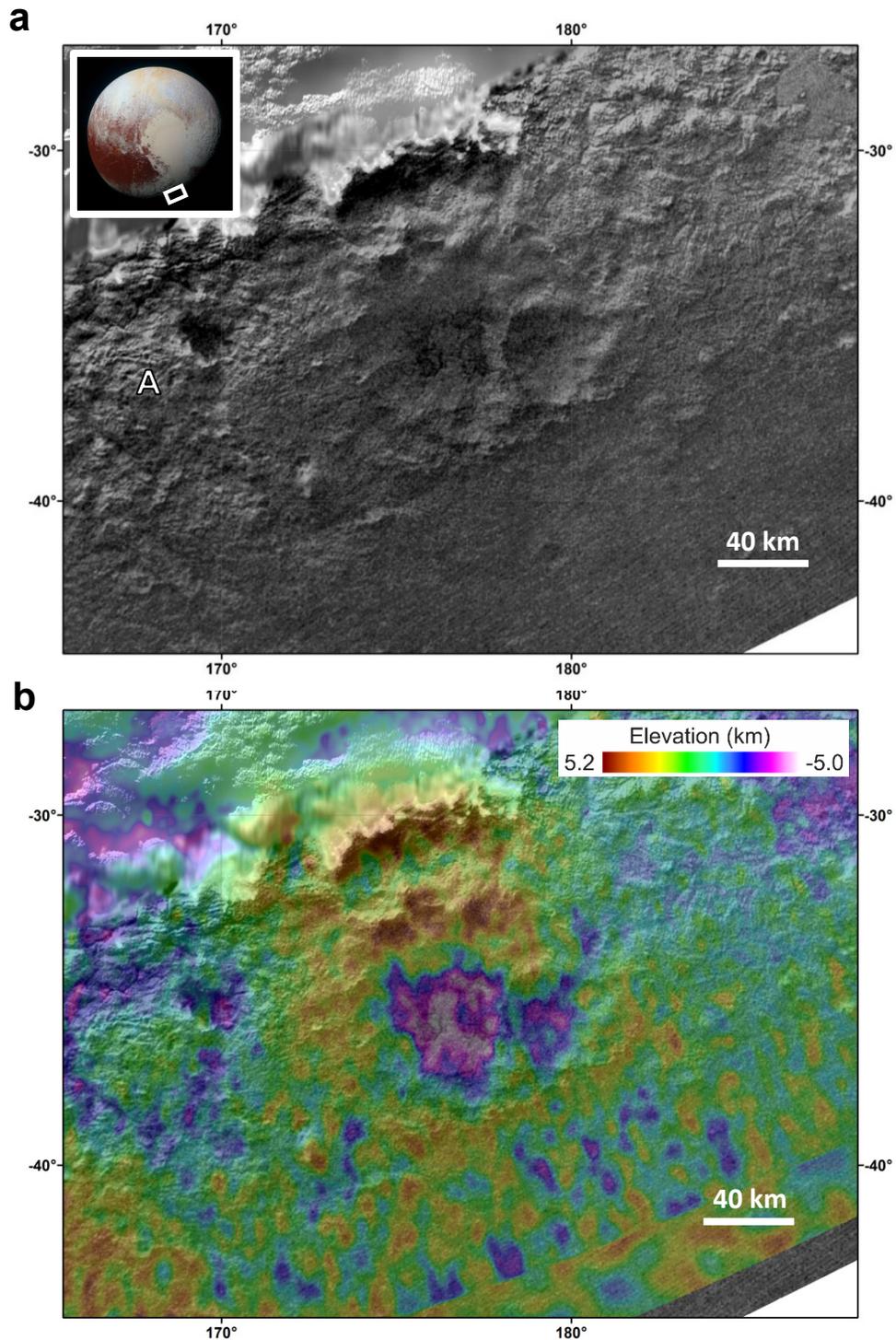

**Supplementary Figure 2 | Piccard Mons and surrounding terrain.** (a) Panchromatic base image (b) topography overlain on basemap. This image is lit by haze light reflected out of Pluto's thin atmosphere.

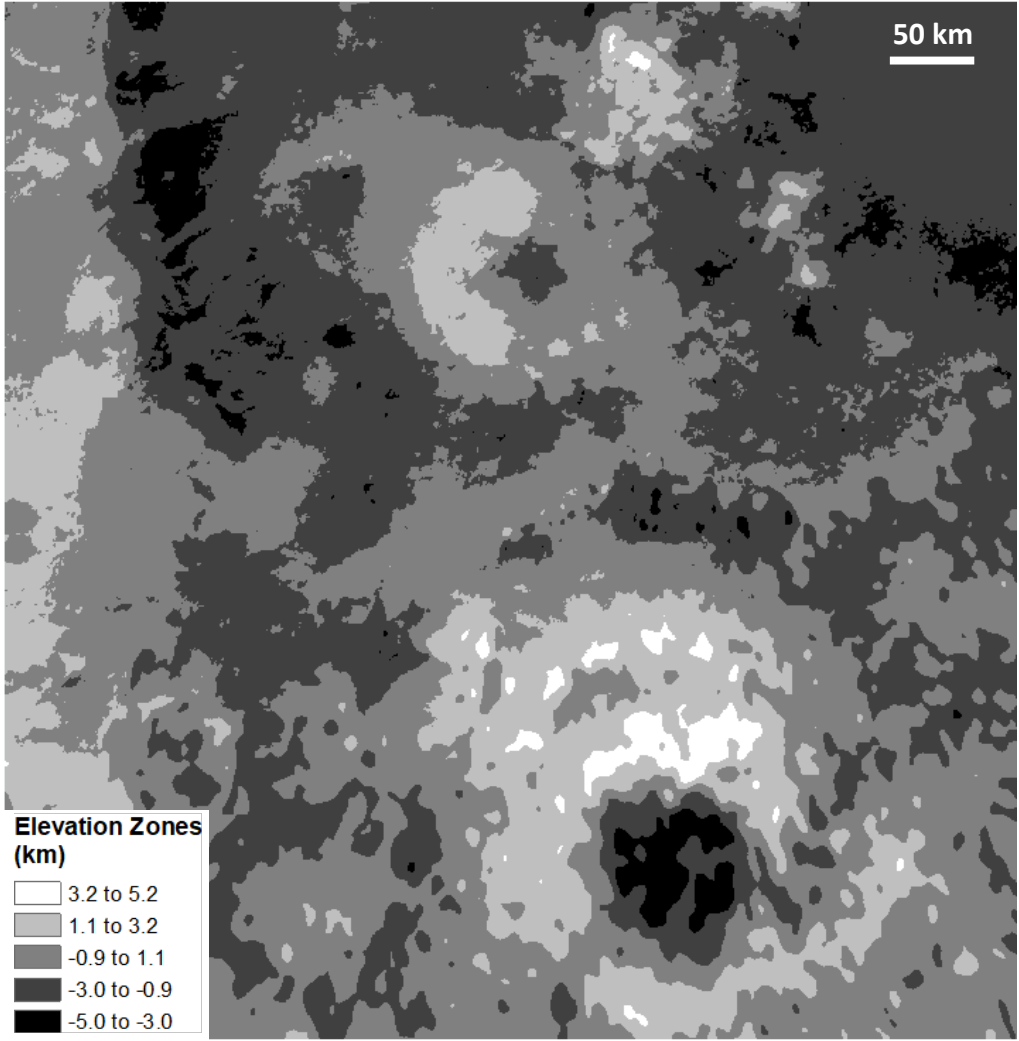

**Supplementary Figure 3 | The topography of the Wright and Piccard regions** here is split into five equal-sized height contours where white is the highest and black is the lowest. This view shows the gross structure of the region and how the "summit" regions around Wright and Piccard are not continuous, and also how the large rise in the medial montes region is connected to both Wright and Piccard Montes. The longitude and latitude extents of the figure are ~160–175°E and ~15–40°S.

**Supplementary Note 3: Hummocky/undulating terrain wavelength and topography**

Here we present some additional information about the undulating/hummocky terrain on the flanks of Wright Mons and the surrounding terrain. The topographic profiles show the variety of morphological expressions of the lumpy terrain (Supplementary Figure 4). Some hummocks have a narrower-sharper boundary between themselves and adjacent hummocks, while others have very shallow or gradual transitions with their neighboring features. However, it is still clear that the hummocks are often not simple individual hills, but are more interconnected to their surroundings and many have one or more "sides" that have no distinct boundary. It is possible that some of the sharper transitions represent the toes of lobate flow fronts, but the oblique lighting (<30° elevation angle) strongly affects the appearance of features in this region.

Thus, we have performed mapping on both the panchromatic image base and the highest resolution topographic product that covers the region. The topography used below is a stereo product produced from an ~315 m px$^{-1}$ MVIC scan and an ~240 m px$^{-1}$ LORRI ride-along mosaic with a vertical precision of 90 m[1] (also see methods). The base image used below for "visual mapping" is the ~315 m px$^{-1}$ MVIC scan which has better signal-to-noise and less smear than the higher resolution 240 m px$^{-1}$ LORRI mosaic.

All mapping involves some subjectivity, but here we attempted to characterize the general wavelength or size of the hummocky-like terrain using a simple metric that can be applied fairly consistently to many of the features. Here we measured across the convex surface of the bulbous mounds or flows in cases where at least two sides of the feature appeared to be bounded by troughs (Supplementary Figure 5). Because the lumpy terrain often does not represent individual hills/hummocks (as described above), we do not attempt to map their full circumference at their base, as this would be quite arbitrary and impossible to consistently define in many cases. We generally avoided the heavily shadowed regions, although the boundaries of the hummocks can be revealed to a degree by stretching the image (here we used built-in ArcGIS routines that optimize the image viewing for low-light regions, rather than bright regions). Although the image in Supplementary Figure 5 is stretched to reveal the brighter portions of the scene, some of the lines extend into the shadows because we were able to see the transitions between features there in a different stretch of the image. The feature edges are more well-defined where they are perpendicular to the direction of the incoming light (i.e., feature edges that trend ~NE-SW are more easily seen), thus more features were measured that trend this direction in the panchromatic image.

Both measurement sets, the panchromatic mapping and the topographic mapping, produced similar results in that most features measured have a wavelength of ~6 and 14 km, with a broad peak between 6-12 km. The western side of this region generally has larger hummocks. The topography shows the hummocks on the northern flank of Wright Mons that are hidden by the oblique lighting in the panchromatic view.

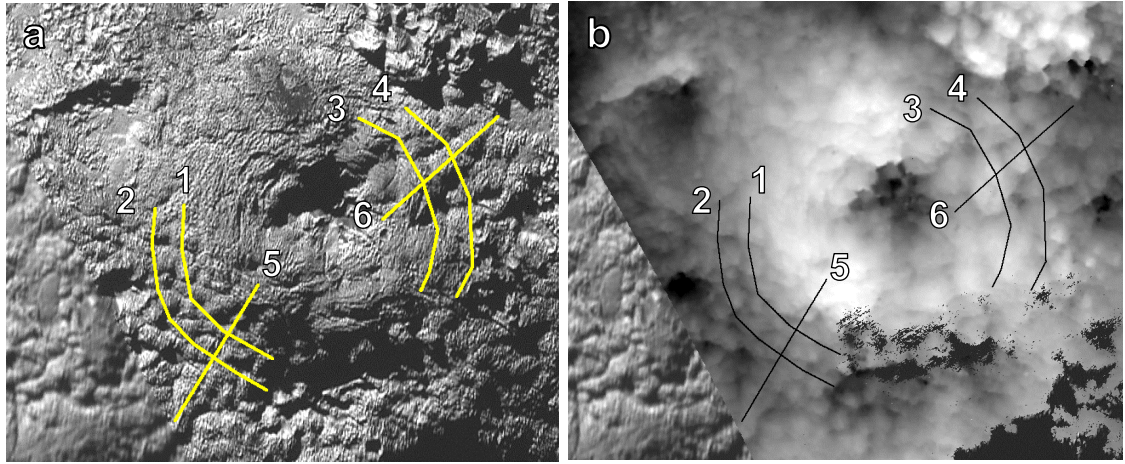
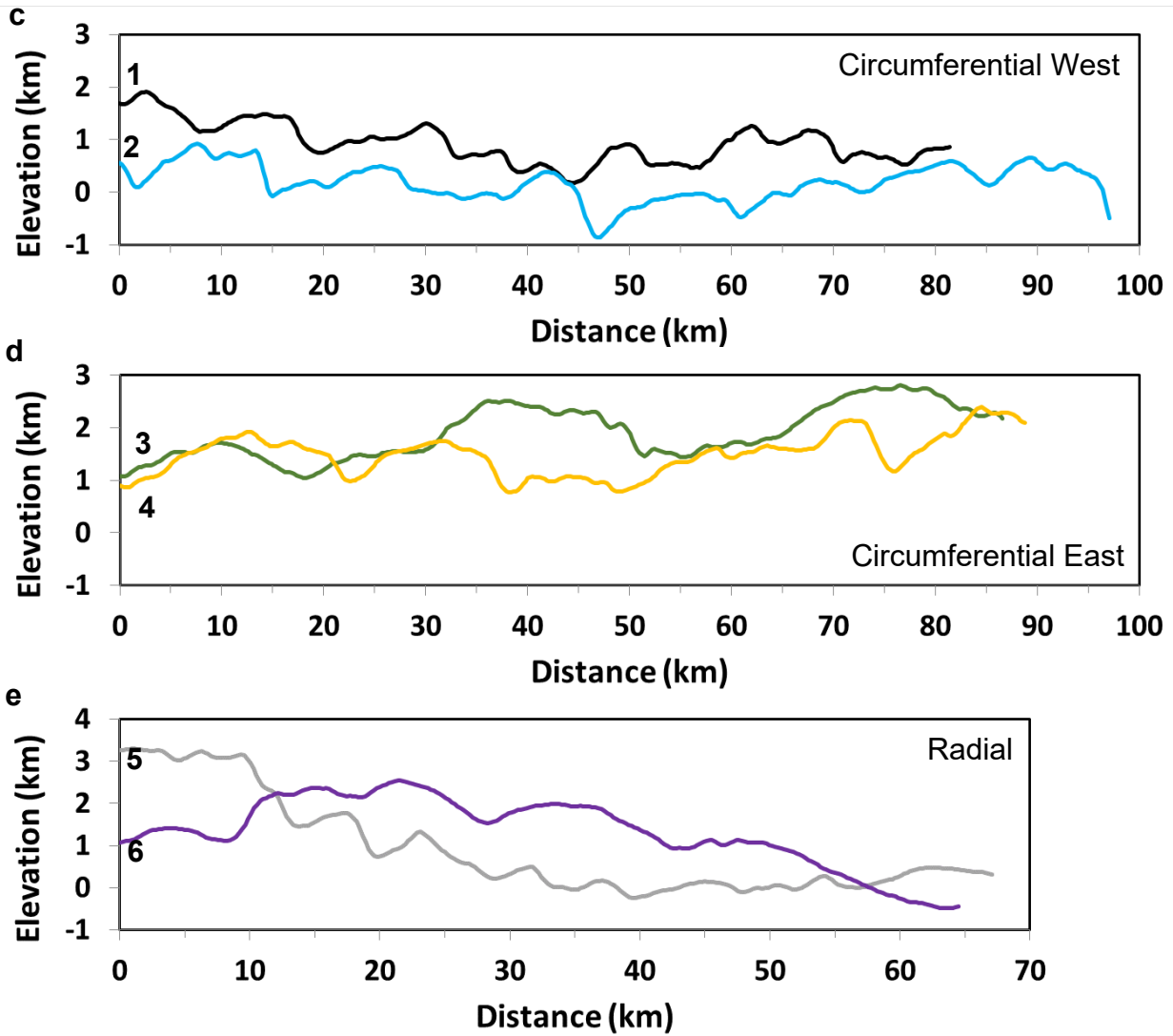

**Supplementary Figure 4 | Topographic profiles of hummocks.** (a) panchromatic image and (b) topography for Wright Mons with topographic profile lines over the undulating/hummocky terrain.

(c-e) Topographic profiles as shown in panels a-b illustrate the varied sizes and shapes of the lumps and depressions between them. Profiles start at the numbered end of each line. The mounds on the eastern side of Wright Mons are larger on average. The longitude and latitude extents of both panels are ~165–177°E and ~17–26°S.

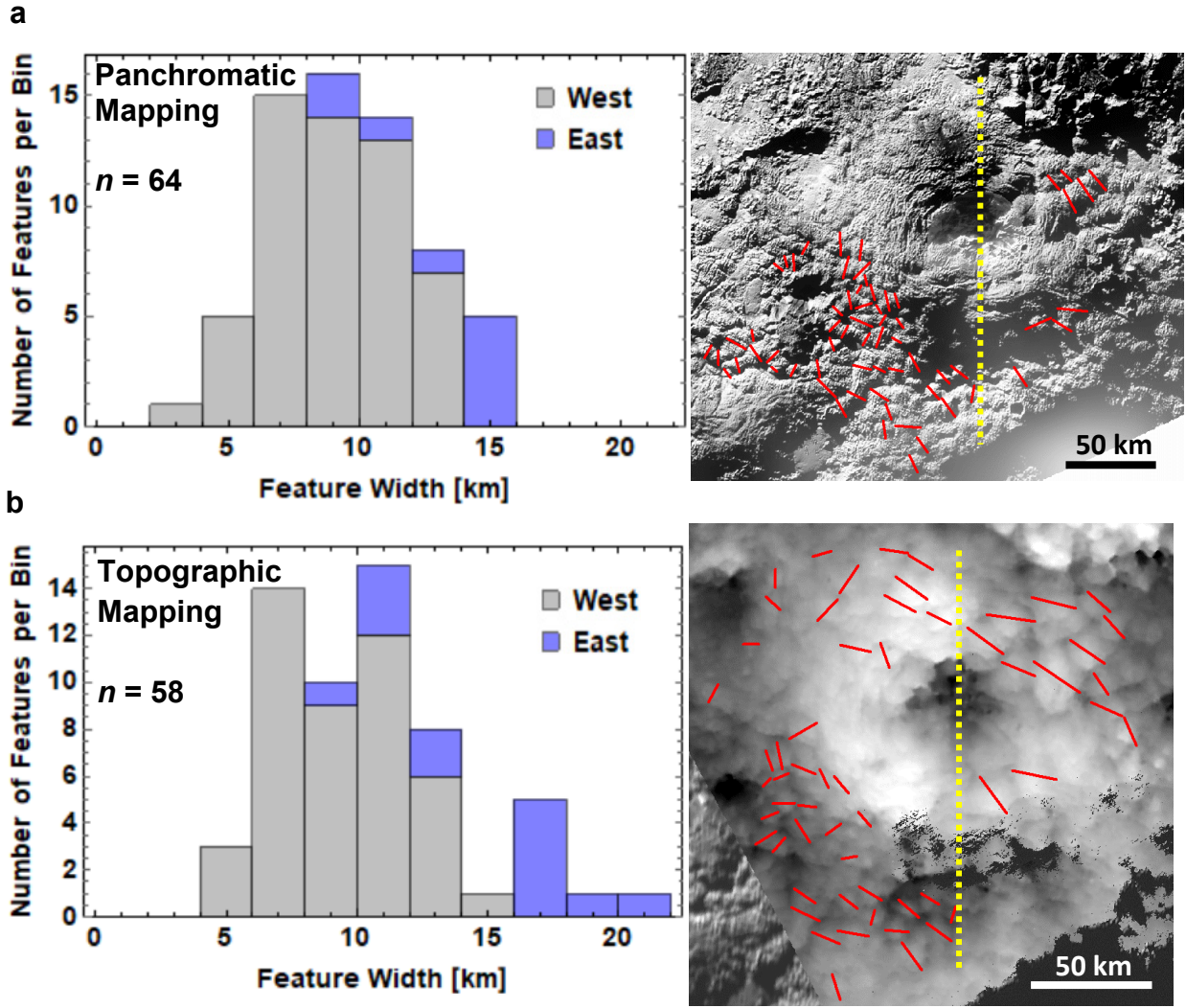

**Supplementary Figure 5 | Hummocky terrain feature measurements. (a)** Results for mapping performed on the panchromatic PEMV_P_MVIC_LORRI_CA dataset (315 m px$^{-1}$). **(b)** Results for mapping performed strictly with the topography. Vertical yellow, dashed line indicates the division between the eastern and western sides of the mapped region. Histogram bin widths were selected via the Sturges method. The longitude and latitude extents of both panels are similar to that of Supplementary Figure 4.

**Supplementary Note 4: Additional close-up views of features in the Wright Mons region**

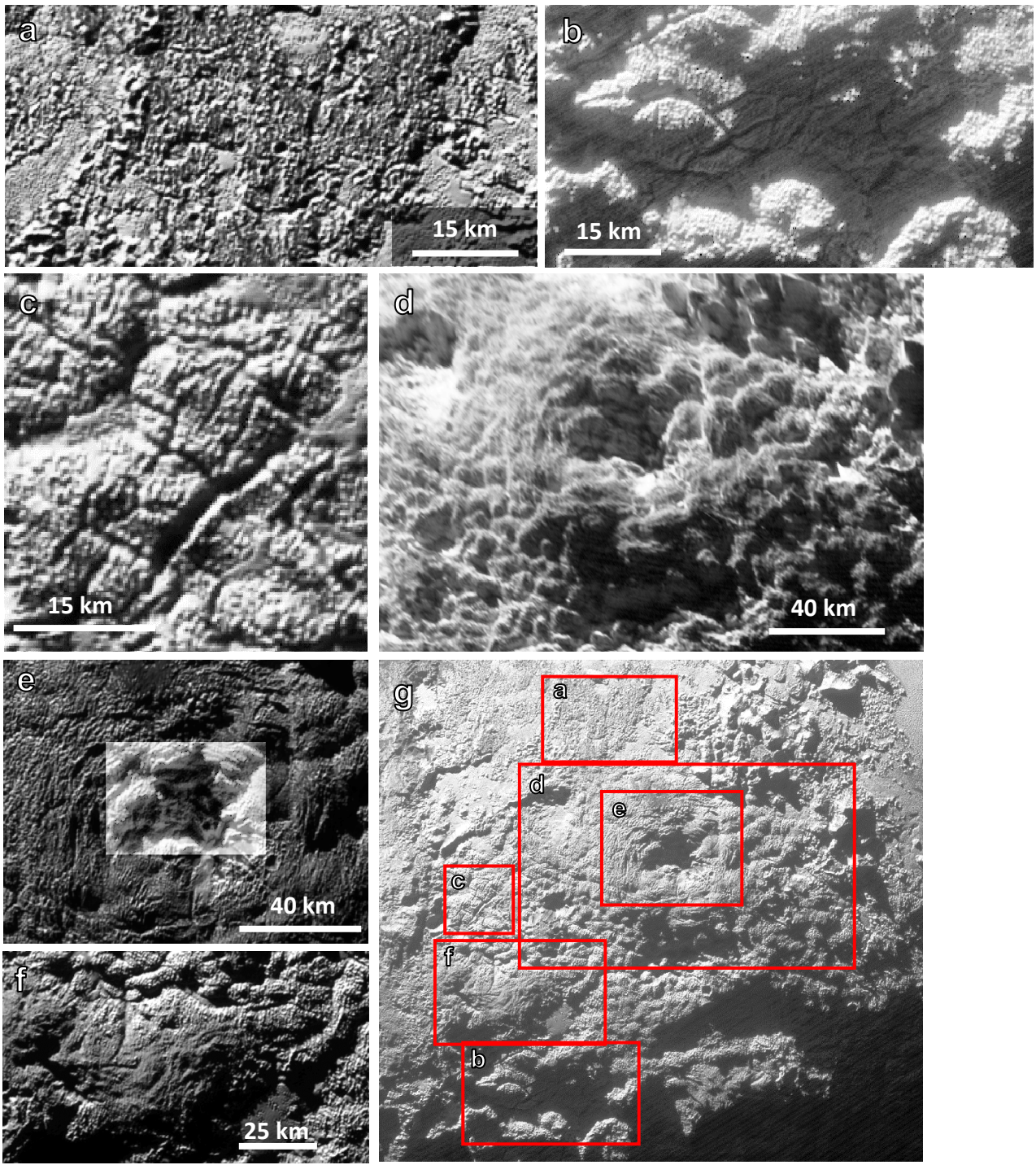

**Supplementary Figure 6 | Additional views and image stretches of Wright Mons and the surrounding terrain.** See text for details. Panels a-c and e-g are from the PEMV_P_MVIC_LORRI_CA observation (315 m/px; see Supplementary Table 1). Panel d is from the HiPhase_HiRes observation (340 m/px). Note that the striping in panel b is from the

instrument scanning, and is not a feature of the surface. Panel g shows the location of the features in this figure and is a subset of Supplementary Figure 1.

**Supplementary Note 5: Comparison with Terrestrial and Martian Volcanic Profiles**

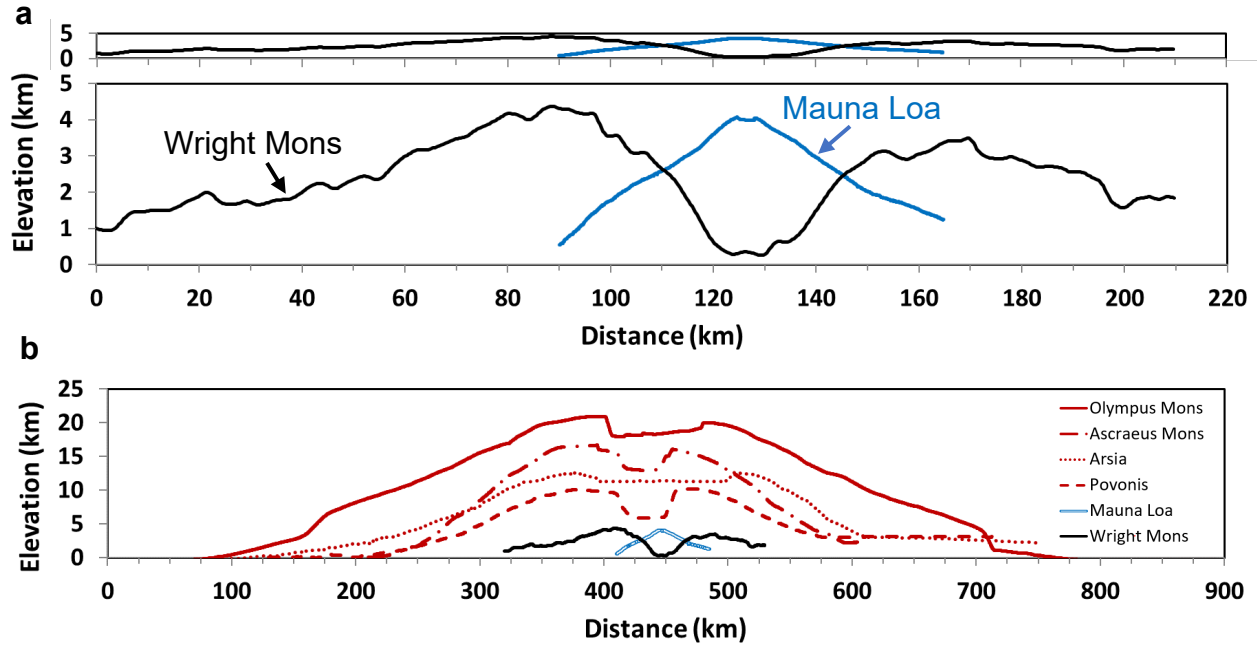

**Supplementary Figure 7 | Topographic profile comparison.** (a) Mauna Loa (subaerial portion only) compared with Wright Mons (~N-S through the feature), upper panel is without vertical exaggeration, and lower panel is shown at 10× vertical exaggeration. Mauna Lau continues for ~6 km below the ocean surface. This figure illustrates how Wright Mons is very dissimilar to Mauna Loa and that if Wright was originally more similar to a shield volcano it would have had to have lost >50% of its volume from the central region in order to attain its current appearance. (b) Martian shield volcanos from the Tharsis region provide additional examples of large volcanos, some with more advanced caldera collapse. Several of these also show signs of later embayment, thus they also may not represent the full original height of the features. The more typical collapse terraces can be seen in the Martian calderas. Note that none of these examples are scaled for gravity, they are shown at their original scales.

**Supplementary Note 6: Comparison with Bladed Terrain on Pluto**

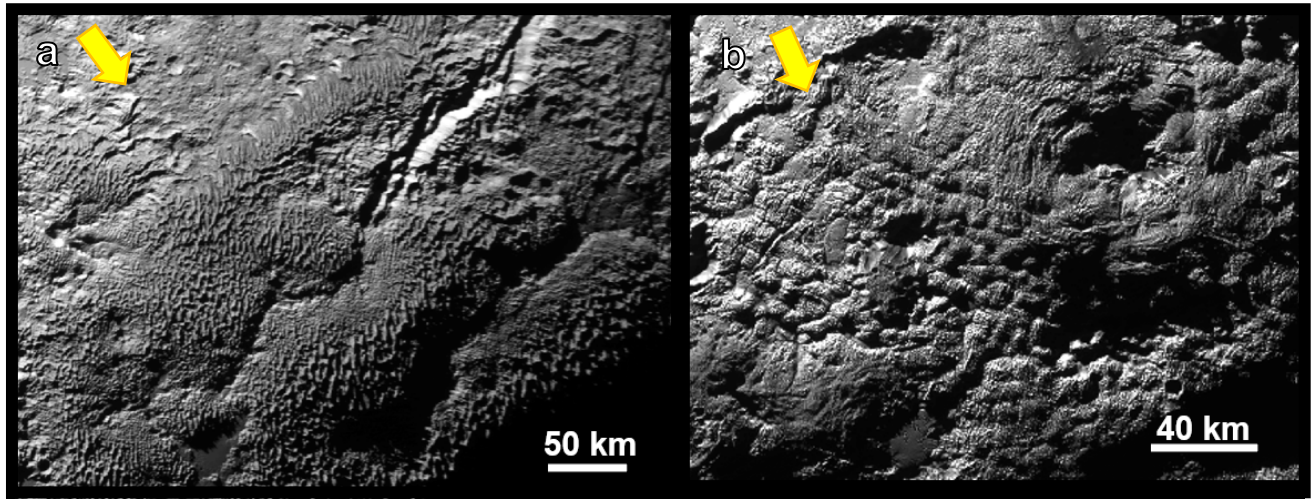

**Supplementary Figure 8 | Comparison of Bladed Terrain and Wright Mons region.** (a) The bladed terrain deposits form one of the highest elevation areas on Pluto (on the far eastern side of the hemisphere observed by New Horizons during closest approach) and are hypothesized to be a sprawling, concentrated deposit of methane ice, with the bladed texture forming due to sublimation of methane [4]. It may also exist in large areas on the "far side" of Pluto that was only observed at low resolution by New Horizons [5]. (b) Although Wright Mons has several different scales and styles of textures, it is not covered by the distinctive blades that characterize the bladed terrain. The longitude and latitude extents are as follows: panels a ~220–243°E and ~10–26°N; panel b ~165–175°E and ~19–26°S. The large arrows in the upper left indicate the approximate direction of the incoming light.

**Supplementary Note 7: Comparison with Funiscular Terrain on Enceladus and Pahoehoe on Earth**

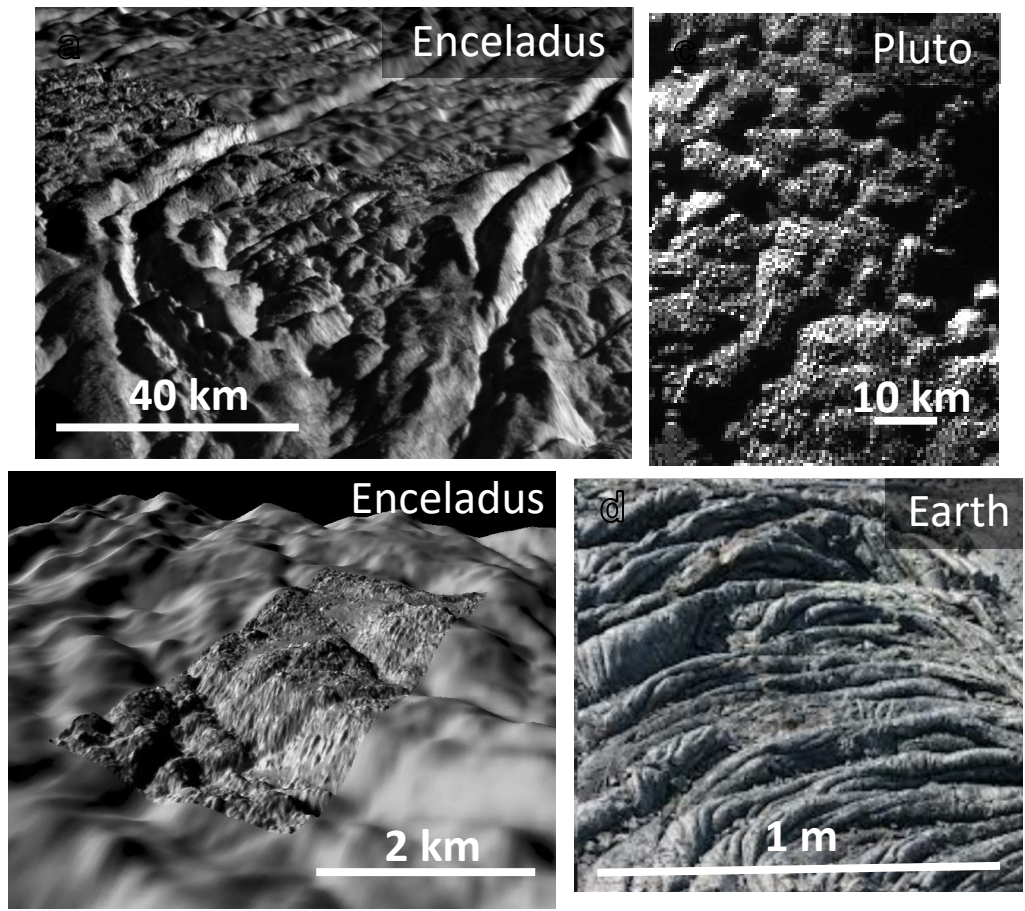

**Supplementary Figure 9 | Funiscular terrain and pahoehoe compared to Pluto.** (a-b) Perspective views of funiscular terrain on Enceladus showing the lumpy material between the tiger stripes has a somewhat consistent size/wavelength (images courtesy NASA/JPL/Space Science Institute/Universities Space Research Association/Lunar & Planetary Institute and can be found at https://photojournal.jpl.nasa.gov/catalog/PIA12208 and https://photojournal.jpl.nasa.gov/catalog/PIA10350. (c) Hummocky/undulatory terrain on Pluto. (d) Pahoehoe lava texture on Earth.

**Supplementary Note 8: Volume Estimate**

The volume of the main topographic rise of the feature outlined as Wright Mons (as shown in Supplementary Figure 10) was measured from the highest elevations on the feature down to an elevation of -1 km in the DEM [1]. An elevation of -1 km is near where the northern base of the large rise transitions to the flatter surrounding terrain, but there is not a clear topographic boundary to the feature on the southern side. This yielded an estimate of ~2.4 x $10^4$ km$^3$ for just this area. Using a lower bound elevation of -0.5 km or -1.5 km (0.5 km higher or lower than our original choice of -1 km) yields volume estimates of ~1.6 x $10^4$ km$^3$ or ~3.3 x $10^4$ km$^3$, respectively (for the same feature outline as shown in Supplementary Figure 10). The uncertainty in the feature outline and the uncertainty in the location of the base of the feature (i.e., how one arbitrarily defines the extent and base location of Wright Mons) is larger than the vertical precision in the DEM (90 m) utilized for this measurement.

Here we provide the volume of this feature as an example. This volume is similar in magnitude to that of the Hawaiian volcano Mauna Loa, estimated to have a volume of ~6-8 x $10^4$ km$^3$ for the total of the subaerial and submerged portions [6]. The resurfacing of the area in general and the creation of the other large rises (e.g., the medial montes region and Piccard Mons) would require a considerable additional volume of material.

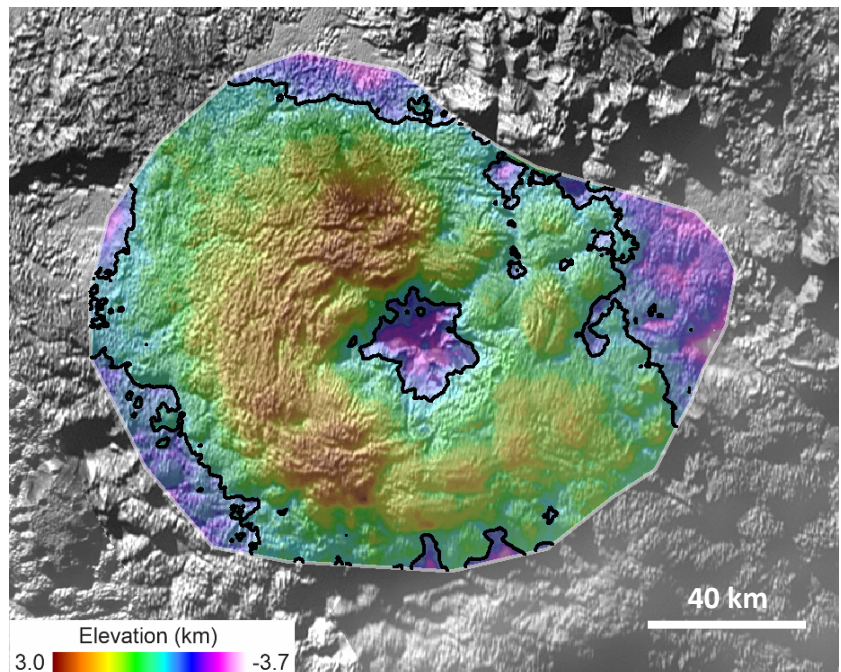

**Supplementary Figure 10 | Wright Mons volume estimate.** The white outline shown here is an approximate boundary for the feature that has been named Wright Mons. The southern part of the feature does not have a clear boundary with the adjacent large rises, but for the sake of measurement we have outlined one region here. The black elevation contour is placed at -1 km in the DEM and was used for the approximate lower boundary of the feature in order to estimate the volume of material needed to create Wright Mons (see text). The longitude and latitude extents of this image are ~167–176°E and ~17–26°S.